\documentclass[aps,prd,preprint,showpacs,showkeys,groupedaddress]{revtex4-1}
\usepackage{latexsym,amssymb,hyperref,graphicx,amsmath,changes}
\usepackage{color}
\usepackage{ulem}   
\begin{document}

\title{Master equation solutions in the linear regime of characteristic formulation of general relativity}

\author{C. E. Cede\~no M.}
\email{eduardo.montana@inpe.br}
\author{J. C. N. de Araujo}
 \email{jcarlos.dearaujo@inpe.br}
\affiliation{INPE - Instituto Nacional de Pesquisas Espaciais}
\date{\today}

\begin{abstract}
From the field equations in the linear regime of the characteristic formulation of general relativity, Bishop, for a Schwarzschild's background, and M\"adler, for a Minkowski's background, were able to show that it is possible to derive a fourth order ordinary differential equation, called master equation, for the $J$ metric variable of the Bondi-Sachs metric. Once $\beta$, another Bondi-Sachs potential, is obtained from the field equations, and $J$ is obtained from the master equation, the other metric variables are solved integrating directly the rest of the field equations. In the past, the master equation was solved for the first multipolar terms, for both the Minkowski's and Schwarzschild's backgrounds.  Also, M\"adler recently reported a generalisation of the exact solutions to the linearised field equations when a Minkowski's background is considered, expressing the master equation family of solutions for the vacuum in terms of Bessel's functions of the first and the second kind. Here, we report new solutions to the master equation for any multipolar moment $l$, with and without matter sources in terms only of the first kind Bessel's functions for the Minkowski, and in terms of the Confluent Heun's functions (Generalised Hypergeometric) for radiative (nonradiative) case in the Schwarzschild's background. We particularize our families of solutions for the known cases for $l=2$ reported previously in the literature and find complete agreement, showing the robustness of our results. 
\end{abstract}
\maketitle
\section{Introduction}
\label{intro}
The characteristic formulation of general relativity offers an attractive point of view to deal with the problem of gravitational wave emission from a source, because this formulation is  based on radiation coordinates. Several complex numerical codes have been developed to treat diverse problems in the nonlinear regime 
\cite{GPW94,BGLW96,BD96,BPR98,BD99,SFMP03,BBSW11,BR14}. However, given the complexity of the  field equations, there are not analytical solutions in the characteristic formulation in this regime. 
\\
In particular in the weak field limit,  this formulation has been used to construct wave extraction algorithms that are applied to obtain the radiation patterns produced in complex numerical simulations of binary systems. Such simulations are usually performed using $3+1$ numerical codes, and then matching algorithms are used in order to make the gravitational wave extraction from some characteristic formulation schemes (see, e.g., \cite{BBSW11,BSWS11}). 
\\ Furthermore, the linear regime has been applied in some interesting situations, despite lacking physical meaning in some cases, such as the equilibrium of a thin shell around a static black hole or in a flat space-time, the motion of a point particle orbiting a Schwarzschild's black hole and the problem of a point particle binary system emitting gravitational radiation \cite{B05,BPR11}. 
\\
In all of these cases the characteristic initial value problem is exchanged by a boundary problem by means of expressing the metric variables as products of their spectral components, using a spin-weighted decomposition, and time oscillatory functions. All these problems deal with matter distributions, in which the field equations could involve terms containing Dirac's delta or Heaviside's functions. The boundary problem is solved imposing regularity in the metric variables at the vertices of the null cones as well as at the null infinity; also, discontinuities in the metric and in their derivatives just across the timelike world tubes which bound the sources are required. This last issue can be done by just following the standard procedures found in the literature, accordingly to the boundary problem to be solved (see, e.g., Bonnor {\cite{B81}}, Georgiou \cite{G92}, Israel \cite{I66}, Choquet-Bruhat \cite{C69}, Taub \cite{T57,T80}, Bishop \cite{B05}.)  
\\ 
We show that it is possible to find analytical solutions to the system of equations for both backgrounds considered (Minkowski and Schwarzschild) using standard methods. In order to do that, it is necessary to transform the problem from partial to ordinary differential equations, through the expansion of the metric variables in spin-weighted spherical harmonics, or in other words, through a multipolar expansion, in which the time dependence is encoded in a periodic function. The substitution of such multipolar expansions into the field equations yield a system of coupled ordinary differential equations. The process to decouple them leads to an equation known as the master equation, which has been solved in the Minkowski's case for the vacuum \cite{M13}, and in presence of a static black-hole for particular values of $l$ \cite{B05}. 
\\
However, it is worth mentioning that it has not been reported so far in the literature any analytical  family of solutions to the master equation with sources in the Minkowski's background. Also, concerning the Schwarzschild's background, there are not any solution for the vacuum nor with any source terms, for any multipolar order. 
\\
We will show in this paper, for the first time in the literature, families of solutions to the master equation, with and without source terms, for arbitrary values of $l$ , for the Minkowski and Schwarzschild's background. 
\\
Here we consider that in both cases studied the sources produce only slight deviations to the background. Consequently, the linear approximation is not taken only for distant points from the sources, but instead it is considered valid for the whole space-time. Within this approximation, we will present the families of solutions to the master equation with and without source terms. In order to do that, some aspects of the characteristic formalism of general relativity and the field equations  are briefly present in Sec. II. The derivation of the master equation is shown in Sec. III; and the solutions to the master equation are shown in Sec. IV. In Sec. V it is shown that these general solutions are reduced to the known solutions for $l=2$. Finally, in Sec. VI we present  some final considerations and conclusions.
\\
\section{The Characteristic and {\it eth} formalisms}
As a starting point, we consider that the space-time is foliated into null cones oriented to the future, which emanate from a central timelike geodesic parametrized by the retarded time $x^1=u$. Consequently, each null cone is labeled by the retarded time. The luminosity distance $x^2=r$ is another parameter measured along the null outgoing rays on the cones. This parameter is chosen in such a way that for $r$ and $u$ constants the spacelike 2-hypersurfaces possess an area of $4\pi r^2$. In addition, the angular coordinates are represented by $x^A$, where $A=3,4$. In these coordinates the Bondi-Sachs metric \cite{BBM62,S62} reads
\begin{align}
\label{bs}
ds^2=&-\left(e^{2\beta}\left(1+\frac{w}{r}\right)-r^2h_{AB}U^AU^B\right)du^2 -2e^{2\beta}dudr 
\nonumber \\& 
-2r^2h_{AB}U^Bdx^A du+r^2h_{AB}dx^Adx^B,
\end{align}
where $\beta$ represents the redshift, $w$ is related to the Newtonian potential, $U^A$ measures the shift of the null cones and $h_{AB}$ represents the metric of the angular manifold. 
\\
The metric of the unit sphere $q_{AB}$ is expressed in terms of dyadic products of the complex vectors $q^A$. These vectors are related to the tangent vectors to the unit sphere which are oriented along the coordinate lines defined by the charts used to make the finite coverage of the sphere \cite{NP66, GMNRS66}. Conventionally, an atlas composed of two stereographic charts, constructed from the poles, are used to cover the unit sphere, and the properties shown here are referred to this particular selection. The metric is then expressed as $q_{AB}=q_{(A}\overline{q}_{B)}$, where the round brackets are denoting symmetrisation with respect to their indices and the overline indicates complex conjugation. In addition, these vectors are null i.e., $q^Aq_A=0$ and satisfy $q_A\overline{q}^A=2$.  They are used to project the angular part of all tensors in the space-time onto the unit sphere. Thus, the angular part of these tensors can be decomposed in spin-weighted scalars with different spin-weights, as described in \cite{NP66,GMNRS66,GPLPW97}. 
\\Consequently, a spin-weighted function $_s\Psi$ with spin-weight $s$ can be constructed from a tensor $\Psi^{a_1\cdots a_n}_{~~~~~~~b_1\cdots b_m}$ in the tangent space to the unit sphere, as
\begin{equation}
_s\Psi=\prod_{i=1}^n \Lambda_{a_i} \prod_{j=1}^m \Lambda^{b_j}\Psi^{a_1\cdots a_n}_{~~~~~~~b_1\cdots b_m}, 
\end{equation}
where $\Lambda_{a_i}$ and $\Lambda^{b_j}$ can take the values $q_{a_i}$ or $\overline{q}_{a_i}$ and $q^{b_j}$ or $\overline{q}^{b_j}$ respectively. The spin-weight $s$ of the functions $_s\Psi$ depends on the number of $q_{a_i}$, $\overline{q}_{a_i}$, $q^{a_i}$ or $\overline{q}^{a_i}$ used to construct them. Thus, if $m,\tilde{s},n,r$ are the number of vectors $q^{a_i},\overline{q}^{a_i},q_{a_i},\overline{q}_{a_i}$ respectively, used to construct the spin-weighted scalar $_s\Psi$, its spin-weight is given by
\begin{equation}
s=2(m+n)-(r+\tilde{s}).
\end{equation}
Hence, the metric for the angular manifold, $h_{AB}$, is decomposed into three spin-weighted scalars $J$, $\overline{J}$ and $K$, with spin-weights $2$, $ -2$, $0$ respectively. Thus,
\begin{equation}
h_{33}=\frac{2(J+\overline{J}+2K)}{(1+|\zeta|^2)^2}, \hspace{0.5cm}h_{34}=-\frac{2i(J-\overline{J})}{(1+|\zeta|^2)^2}, \hspace{0.5cm}h_{44}=-\frac{2(J+\overline{J}-2K)}{(1+|\zeta|^2)^2},
\label{hdefs}
\end{equation}
where $\zeta$ defines the stereographic coordinates, which are related to the spherical coordinates through \begin{equation}
\zeta=\tan\left(\theta/2\right)e^{i\phi}, \hspace{0.5cm}\overline{\zeta}=\tan\left(\theta/2\right)e^{-i\phi};
\end{equation}
and the symbol $|A|$ indicates the norm of the complex scalar $A$. Similarly, the shift vector $U^A$ is decomposed into two spin-weighted scalars $U$ and $\overline{U}$ with spin-weights $1$ and $-1$ respectively, 
\begin{equation}
U=q_AU^A, \hspace{0.5cm}\overline{U}=\overline{q}_AU^A.
\label{Udefs}
\end{equation}
\\
The projections of the covariant derivative related to the unit sphere metric $q_{AB}$ onto the dyads $q^A$ or $q^A$, lead to the differential operators $\eth$ and $\overline{\eth}$. They result in 
\begin{equation}
\eth\ _s\Psi=q^A\ _s\Psi_{,A}+s\Omega\ _s\Psi, \hspace{1.0cm}\overline{\eth}\ _s\Psi=\overline{q}^A\ _s\Psi_{,A}-s\overline{\Omega}\ _s\Psi,
\label{eth_def}
\end{equation}
where the comma indicates partial derivation and $\Omega$ is the contraction $\Omega=-q^Aq^Bq_{A|B}/2$, in which the vertical line was used to represent the covariant derivative with respect to the metric of the unit sphere $q_{AB}$. Notice that  
\eqref{eth_def} highlights the action of these operators on the spin-weighted functions $_s\Psi$, to rise or lower their spin-weights. Through the coordinate transformation between the north and the south charts and from $\eqref{eth_def}$ one obtains
\begin{equation}
\eth\ _s\Psi=\ _{s+1}\Psi, \hspace{1.0cm}\overline{\eth}\ _s\Psi= \ _{s-1}\Psi.
 \label{eth_def1}
\end{equation}
Using \eqref{hdefs} and \eqref{Udefs}, the Bondi-Sachs metric \eqref{bs} can be reexpressed in terms of those spin-weighted scalars. In the linear regime, when the second or higher order terms are disregarded, the metric is reduced to
\begin{align}
\label{bs_lin2}
ds^2&=-du^2-2dudr + \frac{4r^2}{\left(1+|\zeta|^2\right)^2}\left(dq^2+dp^2\right) +\left(\frac{w}{r}+2\beta\right)du^2   \nonumber\\
&-4\beta dudr -\frac{2r^2}{1+|\zeta|^2}du\left((U+\overline{U})dq -i(U-\overline{U})dp \right)
\nonumber \\
&-4ir^2\frac{(J-\overline{J})}{(1+|\zeta|^2)^2}dqdp + \frac{2r^2\left(J+\overline{J}\right)}{\left(1+|\zeta|^2\right)^2}\left(dq^2-dp^2\right),
\end{align} 
which corresponds to a perturbation to the Minkowski metric. 
\\The Einstein's field equations 
\begin{equation}
E_{\mu\nu}=R_{\mu\nu}-8\pi\left(T_{\mu\nu}- g_{\mu\nu}T/2\right)=0,
\end{equation} 
in the characteristic formulation of general relativity \cite{BGLMW97,BGLMW99,RBLTS07} can be written as
\begin{subequations}
	\begin{align}
	& E_{22}=0, \hspace{0.2cm}E_{2A}q^A=0, \hspace{0.2cm} \hspace{0.2cm} E_{AB}h^{AB}=0,\\ 
	& E_{AB}q^{A}q^{B}=0,\\
	& E_{11}=0, \hspace{0.2cm}E_{12}=0,\hspace{0.2cm}\hspace{0.2cm} E_{1A}q^A=0.
	\end{align}
	\label{f_equations}%
\end{subequations}
corresponding to hypersurface, evolution and constrain equations respectively.\\
Explicitly, for the perturbation given in \eqref{bs_lin2} one obtains
\begin{subequations}
	\begin{align}
	& 8 \pi  T_{22}=\frac{4 \beta _{,r}}{r}, \label{field_eq_1}\\
	& 8 \pi  T_{2A} q^A=\frac{\overline{\eth }J_{,r}}{2} -\eth \beta _{,r} +\frac{2 \eth \beta}{r} +\frac{\left(r^4 U_{,r}\right)_{,r}}{2r^2}, \label{field_eq_2}\\
	& 8 \pi  \left(h^{AB} T_{AB}-r^2 T\right) =-2 \eth \overline{\eth }\beta +\frac{\eth^2\overline{J} + \overline{\eth }^2 J}{2} +\frac{\left(r^4\left(\overline{\eth}U+\eth \overline{U}\right)\right)_{,r}}{2r^2}\nonumber\\
	& \hspace{4.0cm}+4 \beta -2 w_{,r}, \label{field_eq_3}\\
	& 8 \pi T_{AB}  q^A q^B=-2 \eth^2 \beta + \left(r^2\eth U \right)_{,r} - \left(r^2 J_{,r}\right)_{,r} +2 r\left(rJ\right)_{,ur},\label{field_eq_4}\\
	& 8 \pi  \left(\frac{T}{2}+T_{11}\right)=\frac{\eth \overline{\eth }w}{2 r^3} + \frac{\eth \overline {\eth }\beta }{r^2} -\frac{\left(\eth \overline{U} + \overline{\eth} U \right)_{,u}}{2} +\frac{w_{,u}}{r^2} +\frac{w_{,rr}}{2 r} \nonumber\\
	&\hspace{2.8cm} -\frac{2 \beta_{,u}}{r}+\frac{2\beta_{,r}}{r}+\beta_{,rr}-2\beta_{,ru}, \label{field_eq_5}
	\\
	& 8 \pi  \left(\frac{T}{2}+T_{12}\right)=
	\frac{\eth \overline{\eth }\beta }{r^2} -\frac{\left(r^2\left(\eth \overline{U} + \overline{\eth   }U\right)\right)_{,r}}{4r^2}+\frac{2\beta_{,r}}{r} +\beta_{,rr}\nonumber\\
	&\hspace{2.8cm}-2\beta_{,ru} +\frac{w_{,rr}}{2 r}, \label{field_eq_6}\\
	& 8 \pi  T_{1A} q^A=\frac{\overline{\eth }J_{,u}}{2} -\frac{\eth^2
		\overline{U} }{4} +\frac{\eth \overline{\eth }U}{4} +\frac{1}{2}\left(\frac{\eth w}{r}\right)_{,r} -\eth \beta _{,u} +\frac{\left(r^4U_{,r}\right)_{,r}}{2r^2}\nonumber\\
	& \hspace{1.8cm}-\frac{r^2 U_{,ur}}{2} +U, \label{field_eq_7}
	\end{align}
	\label{field_eqs}%
\end{subequations}
which were computed previously by Bishop in \cite{B05} for the Schwarzschild's background.\\
\\
Now, given that the eigenfunctions of the $[\eth,\overline{\eth}]$ operators are the spin-weighted spherical harmonics $_sZ_{lm}$, defined in \cite{ZGHLW03} as,
\begin{equation}
_sZ_{lm}=\begin{cases}
\dfrac{i}{\sqrt{2}}\left((-1)^m\ _sY_{lm}+\ _sY_{l\ -m}\right) &\text{for} \hspace{0.5cm} m<0 \\
_sY_{lm}&\text{for} \hspace{0.5cm} m=0 \\
\dfrac{1}{\sqrt{2}}\left(_sY_{lm}+(-1)^m\ _sY_{l\ -m}\right) &\text{for} \hspace{0.5cm} m>0
\end{cases},
\end{equation}
where the spin-weighted spherical harmonics $_sY_{lm}$, are also eigenfunctions of $[\overline{\eth},\eth]$ and are defined in \cite{NP66,GMNRS66,GPLPW97,T07}, as
\begin{equation}
_sY_{lm}=\begin{cases}
\sqrt{\dfrac{(l-s)!}{(l+s)!}}\eth^sY_{lm} & \text{if} \hspace{0.5cm} s\ge 0 \\
(-1)^s\sqrt{\dfrac{(l-s)!}{(l+s)!}}\bar{\eth}^{-s}Y_{lm} & \text{if} \hspace{0.5cm} s< 0
\end{cases}.
\label{sYlm}
\end{equation}
and the fact that spin-weighted spherical harmonics constitutes an orthonormal and complete base of functions, then the metric variables can be expanded in a multipolar series as
\begin{equation}
_sf=\sum_{l=0}^\infty \sum_{m=l}^l \Re\left(f_{lm}e^{i|m|\tilde{\phi}}\right)\ \eth^s\ Z_{lm},
\label{expansion}
\end{equation}
where $_sf=\{\beta,w,J,\overline{J},U,\overline{U}\}$, $Z_{lm}=\ _0Z_{lm}$, $\tilde{\phi}$ is a general function of the retarded time, i.e., $\tilde{\phi}:=\tilde{\phi}(u)$, $f_{lm}$ are the spectral components of the function $_sf$, $m\in \mathbb{Z}$, $m\in [-l,l]$ and $l\ge 0$ indicating the multipolar order. 
\\ Notice that in \eqref{expansion} the spin-weight of the function $_sf$ is contained in $\eth^s Z_{lm}$. Therefore, substituting \eqref{expansion} into the field equations \eqref{field_eqs} one obtains ordinary differential equations for their spectral components, in which the spin-weighted factors have been eliminated, namely  
\begin{subequations}
	\begin{align}
	&\beta_{lm,r}= 2\pi \int_{\Omega}d\Omega\ \overline{Z}_{lm}\int_{0}^{2\pi}d\tilde\phi \ e^{-i|m|\tilde\phi} r T_{22} 
	\label{field_eq_s1} ,
	\end{align}
	\begin{align}
	& -\frac{(l+2)(l-1)J_{lm,r}}{2} -\beta _{lm,r} +\frac{2 \beta_{lm}}{r} +\frac{\left(r^4 U_{lm,r}\right)_{,r}}{2r^2} 
	\nonumber\\
	&
	= \frac{8 \pi}{\sqrt{l(l+1)}} \int_{\Omega} d\Omega \ \overline{Z}_{lm}\int_{0}^{2\pi} d\tilde\phi \ e^{-i|m|\tilde{\phi}} T_{2A} q^A, 
	\label{field_eq_s2}
	\end{align}
	\begin{align}
	& 2l(l+1) \beta_{lm} +(l-1)l(l+1)(l+2)J_{lm}  +\frac{l(l+1)\left(r^4\left(U_{lm}\right)\right)_{,r}}{r^2}\nonumber\\
	& + 4 \beta_{lm} -2 w_{lm,r} = 8 \pi \int_{\Omega}d\Omega \ \overline{Z}_{lm}\int_{0}^{2\pi}d\tilde\phi\ e^{-i|m|\tilde{\phi}}\left(h^{AB} T_{AB}-r^2 T\right) , \label{field_eq_s3}
	\end{align}
	\begin{align}
	&-2 \beta_{lm} + \left(r^2 U_{lm} \right)_{,r} - \left(r^2 J_{lm,r}\right)_{,r} +2 i|m| r\dot{\tilde{\phi}} \left(rJ_{lm}\right)_{,r} \nonumber\\
	&= \frac{8 \pi}{\sqrt{(l-1)l(l+1)(l+2)}} \int_{\Omega}d\Omega \ \overline{Z}_{lm} \int_{0}^{2\pi}d\tilde{\phi}\ e^{-i|m|\tilde{\phi}}T_{AB}  q^A q^B,
	\label{field_eq_s4}
	\end{align}
	\begin{align}
	& -\frac{l(l+1) w_{lm}}{2 r^3} - \frac{l(l+1)\beta_{lm} }{r^2} +i|m|l(l+1)\dot{\tilde{\phi}} U_{lm}  +\frac{i|m|\dot{\tilde{\phi}} w_{lm}}{r^2}   \nonumber\\
	&+\frac{w_{lm,rr}}{2 r} -\frac{2 i|m|\dot{\tilde{\phi}} \beta_{lm}}{r}+\frac{2\beta_{lm,r}}{r} + \beta_{lm,rr}-2\dot{\tilde{\phi}}\beta_{lm,r}\nonumber\\
	&=8 \pi \int_{\Omega}d\Omega \ \overline{Z}_{lm} \int_{0}^{2\pi}d\tilde{\phi}\ e^{-i|m|\tilde{\phi}} \left(\frac{T}{2}+T_{11}\right), \label{field_eq_s5}\\
	& -\frac{l(l+1)\beta_{lm} }{r^2} +\frac{l(l+1)\left(r^2 U_{lm} \right)_{,r}}{2r^2} +\frac{w_{lm,rr}}{2 r} \nonumber\\
	&=8 \pi \int_{\Omega}d\Omega \ \overline{Z}_{lm} \int_{0}^{2\pi}d\tilde{\phi}\ e^{-i|m|\tilde{\phi}} \left(\frac{T}{2}+T_{12}\right), \label{field_eq_s6}
	\end{align}
	\begin{align}
	& -\frac{i|m|(l+2)(l-1)J_{lm}\dot{\tilde{\phi}}}{2} +\frac{1}{2}\left(\frac{w_{lm}}{r}\right)_{,r} -i|m|\dot{\tilde{\phi}} \beta _{lm} +\frac{\left(r^4U_{lm,r}\right)_{,r}}{2r^2}\nonumber\\
	& -\frac{i|m|r^2\dot{\tilde{\phi}}}{2}U_{lm,r} +U_{lm} =\frac{8 \pi}{\sqrt{l(l+1)}}  \int_{\Omega}d\Omega \ \overline{Z}_{lm} \int_{0}^{2\pi}d\tilde{\phi}\ e^{-i|m|\tilde{\phi}} T_{1A} q^A, \label{field_eq_s7}
	\end{align}
	\label{field_eq_s}
\end{subequations}
This system of coupled ordinary equations is separable through a simple procedure, as we will show in the next section. Notice that an alternative procedure is presented by M\"adler in \cite{M13}.
\section{The master equation}
Making the change of variable $x=r^{-1}$, the field equations \eqref{field_eq_s1} - \eqref{field_eq_s4} become 
\begin{subequations}	
	\begin{align}
	&\beta_{lm,x}= -x^2A_{lm},  
	\label{field_eq_x1}\\
	& (l+2)(l-1)xJ_{lm,x} +2x\beta _{lm,x} +4 \beta_{lm} -2U_{lm,x}+xU_{lm,xx} = B_{lm}, 
	\label{field_eq_x2}\\
	& - 2x^3 J_{lm,xx}  -4 i|m|\dot{\tilde{\phi}} x J_{lm,x} +4i|m| \dot{\tilde{\phi}} J_{lm} +4 U_{lm} - 2xU_{lm,x}-4 x\beta_{lm} 
	\nonumber\\
	&
	= 2x D_{lm},
	\label{field_eq_x4}
	\end{align}
\end{subequations}
where the source terms $A_{lm}:=A_{lm}(x)$, $B_{lm}:=B_{lm}(x)$ and $D_{lm}:=D_{lm}(x)$ are explicitly defined,
\begin{subequations}
	\begin{align}
	& A_{lm}=2\pi \int_{\Omega}d\Omega\ \overline{Z}_{lm}\int_{0}^{2\pi}d\tilde\phi \ e^{-i|m|\tilde\phi} x T_{22},\\
	& B_{lm}=\frac{16 \pi}{\sqrt{l(l+1)}} \int_{\Omega} d\Omega \ \overline{Z}_{lm}\int_{0}^{2\pi} d\tilde\phi \ e^{-i|m|\tilde{\phi}} x T_{2A} q^A,\\
	& D_{lm}= \frac{8 \pi}{\sqrt{(l-1)l(l+1)(l+2)}} \int_{\Omega}d\Omega \ \overline{Z}_{lm} \int_{0}^{2\pi}d\tilde{\phi}\ e^{-i|m|\tilde{\phi}}T_{AB}  q^A q^B.
	\end{align}
\end{subequations}
In addition, solving \eqref{field_eq_x2} for $4x\beta_{lm}$ 
and substituting it into \eqref{field_eq_x4}, one obtains
\begin{align}
& - 2x^3 J_{lm,xx}  -4 i|m| \dot{\tilde{\phi}}xJ_{lm,x} + x^2(l+2)(l-1)J_{lm,x} +4i|m| \dot{\tilde{\phi}} J_{lm}   \nonumber \\
&  +x^2U_{lm,xx} - 4xU_{lm,x} +4 U_{lm} +2x^2\beta _{lm,x}  = x (2D_{lm}+ B_{lm}).
\label{field_eq_x5}
\end{align}
Thus, the derivative of \eqref{field_eq_x5} with respect to $x$ yields  a third order differential equation for $J_{lm}$, i.e., 
\begin{align}
& - 2x^3 J_{lm,xxx} - 6x^2 J_{lm,xx}  -4 i|m| \dot{\tilde{\phi}}xJ_{lm,xx} + x^2(l+2)(l-1)J_{lm,xx} \nonumber\\
&+ 2x(l+2)(l-1) J_{lm,x}   +x^2U_{lm,xxx}  - 2xU_{lm,xx}   \nonumber \\
&  +4x\beta _{lm,x} +2x^2\beta _{lm,xx} = (2D_{lm}+ B_{lm}) + x (2D_{lm,x}+ B_{lm,x}).
\label{field_eq_x6}
\end{align}
After this, notice that it is possible to obtain $x^2U_{lm,xxx}$ just deriving \eqref{field_eq_x2} with respect to $x$, 
\begin{align}
&  x^2U_{lm,xxx} = -x^2(l+2)(l-1)J_{lm,xx} -x(l+2)(l-1)J_{lm,x}  +xU_{lm,xx}  \nonumber\\
&  -6x\beta _{lm,x} -2x^2\beta _{lm,xx} + x B_{lm,x}.
\label{field_eq_x7}
\end{align}
Then, substituting it in \eqref{field_eq_x6} and simplifying one obtains
\begin{align}
& - 2x^3 J_{lm,xxx} - 6x^2 J_{lm,xx}  -4 i|m| \dot{\tilde{\phi}}xJ_{lm,xx} + x(l+2)(l-1) J_{lm,x} \nonumber\\
&  -xU_{lm,xx}  -2x\beta _{lm,x} = 2 x D_{lm,x}+B_{lm}+2 D_{lm}.
\label{field_eq_x8}
\end{align}
Making the derivative of \eqref{field_eq_x8} with respect to $x$, and substituting  $xU_{xxx}$ from \eqref{field_eq_x7} one  finds a fourth order differential equation for $J_{lm}$, namely
\begin{align}
&  - 2x^4 J_{lm,xxxx} - 12x^3 J_{lm,xxx} - 12x^2 J_{lm,xx}   -4i|m|  \dot{\tilde{\phi}}xJ_{lm,xx} -4i|m|  \dot{\tilde{\phi}} x^2 J_{lm,xxx} \nonumber\\
&+ 2x(l+2)(l-1) J_{lm,x} + 2x^2(l+2)(l-1) J_{lm,xx} +4x\beta _{lm,x} -2x U_{lm,xx}  \nonumber\\
&= 2 x B_{lm,x}+2 x^2 D_{lm,xx}+4 x D_{lm,x}.
\label{field_eq_x10}
\end{align}
Finally, solving \eqref{field_eq_x8} for $U_{lm,xx}$ and substituting into \eqref{field_eq_x10}, a differential equation containing only $J_{lm}$ with source terms is obtained, namely
\begin{align}
&  - 2x^4 J_{lm,xxxx} - 4x^2\left(2x  + i|m|\dot{\tilde{\phi}}\right)J_{lm,xxx} \nonumber\\ &+2x\left(2i|m|  \dot{\tilde{\phi}} + x(l+2)(l-1) \right)J_{lm,xx}  
   = H_{lm}(x),
\label{field_eq_x11}
\end{align}
where 
\begin{align}
H_{lm}(x)=2 x B_{lm,x}+2 x^2 D_{lm,xx}-8 x \beta_{lm,x}-2 B_{lm}-4 D_{lm}
\end{align}
represents the source terms.\\
In order to reduce the order of this differential equation, one  defines \\  $\tilde{J}_{lm}=J_{lm,xx}$, thus,
\begin{align}
&  - 2x^4 \tilde{J}_{lm,xx} - 4x^2\left(2x  + i|m|\dot{\tilde{\phi}}\right)\tilde{J}_{lm,x} +2x\left(2i|m|  \dot{\tilde{\phi}} + x(l+2)(l-1) \right)\tilde{J}_{lm}  
   = H_{lm}.
\label{field_eq_x12}
\end{align}
For the vacuum, this differential equation turns homogeneous, i.e.,  $H_{lm}=0$, and hence \eqref{field_eq_x12} is reduced to the master equation presented by M\"adler in \cite{M13}
\begin{align}
- x^3 \tilde{J}_{lm,xx} - 2x\left(2x  + i|m|\dot{\tilde{\phi}}\right)\tilde{J}_{lm,x} +\left(2i|m|  \dot{\tilde{\phi}} + x(l+2)(l-1) \right)\tilde{J}_{lm}  
= 0.
\label{field_eq_x13}
\end{align}
Making $l=2$, this master equation reduces to those presented previously in \cite{B05} for the Minkowski's background i.e.,
\begin{align*}
- x^3 \tilde{J}_{lm,xx} - 2x\left(2x  + i|m|\dot{\tilde{\phi}}\right)\tilde{J}_{lm,x} +2\left(i|m|  \dot{\tilde{\phi}} + 2x \right)\tilde{J}_{lm}  = 0.
\end{align*}
The derivation of the master equation for the Schwarzschild's background follows the same scheme. In this case the master equation is given by
\begin{align}
& J_{lm,xxxx}x^4 (2 M x-1) + J_{lm,xxx} \left(2 x^3 (7 M x-2)-2 i x^2 \dot{\tilde{\phi }}
\left| m\right| \right)\nonumber\\
&+J_{lm,xx} \left(2 i x \dot{\tilde{\phi }} \left| m\right| +(l-1) (l+2) x^2+16 M
x^3\right)=G_{lm}(x),
\label{meqsch1}
\end{align}
where $M$ is the mass of the central static black-hole and $G_{lm}(x)$ represents the source term, which is given by
\begin{equation}
G_{lm}(x)=\frac{H_{lm}(x)}{2}.
\end{equation}
It is important to observe that $M=0$ effectively reduces \eqref{meqsch1} to \eqref{field_eq_x11}. 
\\
Defining $\tilde{J}_{lm}=J_{lm,xx}$, the order of the differential equation \eqref{meqsch1} is reduced, namely 
\begin{align}
& \tilde{J}_{lm,xx}x^4 (2 M x-1) + \tilde{J}_{lm,x} \left(2 x^3 (7 M x-2)-2 i x^2 \dot{\tilde{\phi }}
\left| m\right| \right)\nonumber\\
&+\tilde{J}_{lm} \left(2 i x \dot{\tilde{\phi }} \left| m\right| +(l-1) (l+2) x^2+16 M
x^3\right)=G_{lm}(x).
\label{meqsch2}
\end{align}
\section{Families of solutions to the master equation}
Now, the families of solutions to the master equations \eqref{field_eq_x11} and \eqref{meqsch1} associated with the linear approximation in the Minkowski and the Schwarzschild's space-times are explicitly shown. 
\\
To proceed, consider that $l$ is an integer and greater than or equal to zero, i.e., $l\ge 0$, the constants of integration $C_i$ are complexes $C_i\in \mathbb{C}$, $i=1..4$, and arabic lower case letters represent real constants, i.e., $a,b,c,d,e,f,\cdots \in \mathbb{R}$
\\ It is worth stressing that the applicability of the present work has some limitations,
since in the context of the characteristic formulation the matter fields must be known {\it a
priori} throughout the spacetime.
\subsection{The Minkowski's background}
First, let us consider the most simple case corresponding to the nonradiative, $m=0$, Minkowski's master equation without sources \eqref{field_eq_x13}. Assuming the ansatz $J_{lm}=x^k$, we obtain immediately 
\begin{equation*}
(k-l+1) (k+l+2)=0
\end{equation*}
whose roots leave us to the general family of solutions, 
\begin{equation}
\tilde{J}_{l0}(x)= C_1 x^{l-1}+C_2 x^{-(l+2)}.
\label{Msol1}
\end{equation}
Thus, integrating the last equation two times and rearranging the constants one obtains families of solutions to \eqref{field_eq_x11} of four parameters for the vacuum,
\begin{equation}
J_{l0}(x)= C_1 x^{l+1}+C_2 x^{-l}+C_3x+C_4.
\label{Msol2}
\end{equation}
When the source term is not null, we find that the nonradiative family of solutions, $m=0$, to the inhomogeneous equation \eqref{field_eq_x12} reads
\begin{align}
\tilde{J}_{l0}(x)=& C_1 x^{l-1} + C_2 x^{-(l+2)}+ x^{-(l+2)} \int_a^x dy\, \frac{H(y) y^{l-1}}{2 l+1} -x^{l-1} \int_b^x dy\, 
\frac{H(y) y^{-(l+2)}}{2 l+1}, 
\label{Msol3}
\end{align}
where $a$ and $b$ are real constants. Therefore, integrating two times with respect to $x$ and rearranging the constants we find the family of solutions to the inhomogeneous master equation \eqref{field_eq_x11}, for $m=0$,
\begin{align}
J_{l0}(x)=& C_1 x^{l+1}+ C_2 x^{-l}+ C_3x +C_4
+ \int_a^x dv\, \int_{b}^{v}dw \, w^{-(l+2)} \int_c^w dy\, \frac{H(y) y^{l-1}}{4 l+2}
\nonumber\\
&  -\int_d^xdv\, \int_{e}^{v} dw\, w^{l-1} \int_f^w dy\, \frac{H(y) y^{-(l+2)}}{4 l+2},
\label{Msol4}
\end{align}
where it is clear that the analyticity of the solutions depends on the existence and analyticity of the integrals. If the source term is disregarded, then \eqref{Msol4} is reduced immediately to \eqref{Msol2}.
\\Now, we will consider the case for a radiative family of solutions, $m\ne 0, \ |m|\le l$ for $l> 0$, without source term. In this case \eqref{field_eq_x13} becomes a Bessel's type differential equation. M\"adler \cite{M13} previously shows that the general solutions to this master equation can be expressed as a linear combination of the first and second kind spherical Bessel's functions. We find here that the family of solutions to the master equation  \eqref{field_eq_x13} can be expressed in terms only of the first kind Bessel's functions, as
\begin{align}
\tilde{J}_{lm}&=\frac{C_1 2^{\frac{1}{2}-2 l} z^{3/2} e^{\frac{1}{2} i (\pi  l+2 z)} \Gamma \left(\frac{1}{2}-l\right) \left(K
	J_{-l-\frac{1}{2}}+L J_{\frac{1}{2}-l}\right)}{(l-1) l}\nonumber\\
&+\frac{i C_2 2^{2 l+\frac{5}{2}} z^{3/2} e^{i
		z-\frac{i \pi  l}{2}} \Gamma \left(l+\frac{3}{2}\right) \left(K J_{l+\frac{1}{2}}+L
	J_{l-\frac{1}{2}}\right)}{(l+1) (l+2)},
\label{mes2}
\end{align}
where the argument of the first kind Bessel's functions $J_{n}$ are referred to $z$, which is defined as
\begin{equation}
z=\dfrac{|m|\dot{\tilde{\phi}}}{x},
\label{zdef}
\end{equation}
and the coefficients $K$, $L$ and $S$ are,
\begin{subequations}
	\begin{align}
    K&=-i (l(l-1)+2iz) -2z(l-i z),\\
	L&=-2 z (z-i),\\
	S&=l(l-1)+2iz.
	\end{align}
\end{subequations}
Integrating two times \eqref{mes2}, and rearranging the constants we find the family of solutions that satisfies \eqref{field_eq_x11}, i.e.,
\begin{align}
J_{lm}=&-\frac{iC_1 2^{\frac{1}{2}-2 l} \dot{\tilde{\phi}} ^2 \left| m\right| ^2 z^{-1/2}e^{\frac{1}{2} i (\pi  l+2 z)} \Gamma
	\left(\frac{1}{2}-l\right) \left(-2 z J_{\frac{1}{2}-l}+\overline{S}
	J_{-l-\frac{1}{2}}\right)}{l^2 \left(l^2-1\right)}\nonumber\\
&-\frac{C_22^{2 l+\frac{5}{2}} \dot{\tilde{\phi}} ^2 \left| m\right| ^2 z^{-1/2}e^{-\frac{1}{2} i (\pi  l-2 z)} \Gamma
	\left(l+\frac{3}{2}\right) \left(2 z J_{l-\frac{1}{2}}+\overline{S}
	J_{l+\frac{1}{2}}\right)}{l (l+1)^2 (l+2) }\nonumber\\
&+C_3+C_4\frac{\dot{\tilde{\phi}} |m|}{z}.
\end{align}
When matter is considered, we found that the family of solutions to \eqref{field_eq_x13} becomes,
\begin{align}
\tilde{J}_{lm}=&\frac{2^{\frac{1}{2}-2 l} z^{3/2} \left(C_1+D_1\right) e^{\frac{i \pi  l}{2}+i z} \Gamma
	\left(\frac{1}{2}-l\right) \left(K J_{-l-\frac{1}{2}}+L J_{\frac{1}{2}-l}\right)}{(l-1) l}\nonumber\\
&+\frac{i 2^{2
		l+\frac{5}{2}} z^{3/2} \left(C_2+D_2\right) e^{i z-\frac{i \pi  l}{2}} \Gamma \left(l+\frac{3}{2}\right)
	\left(K J_{l+\frac{1}{2}}+L J_{l-\frac{1}{2}}\right)}{(l+1) (l+2)},
\label{mes3}
\end{align}
where the coefficients $K$ and $L$ were defined above, and the terms representing sources are
\begin{subequations}	
\begin{align}
D_1=-&\int_{|m|\dot{\tilde{\phi}}}^{|m|\dot{\tilde{\phi}}/z}d\tilde{z}\, \frac{2^{2 l-\frac{5}{2}} \tilde{z}^{-1/2} e^{-\frac{1}{2} i (\pi  l+2 \tilde{z})} \Gamma \left(l+\frac{1}{2}\right)  \left(K J_{l+\frac{1}{2}}-L
	J_{l-\frac{1}{2}}\right)}{(l+1) (l+2) \dot{\tilde{\phi}} ^2 \left| m\right| ^2}H\left(\frac{\dot{\tilde{\phi}}  \left| m\right| }{\tilde{z}}\right),
\end{align}
and
\begin{align}
D_2=&-i\int_{|m|\dot{\tilde{\phi}}}^{|m|\dot{\tilde{\phi}}/z}d\tilde{z}\, \frac{ 2^{-2 l-\frac{9}{2}}\tilde{z}^{-1/2} e^{\frac{1}{2} i (\pi  l-2 \tilde{z})} \Gamma \left(-l-\frac{1}{2}\right)  \left(K J_{-l-\frac{1}{2}}+L
	J_{\frac{1}{2}-l}\right)}{(l-1) l \dot{\tilde{\phi}} ^2  \left| m\right| ^2} H\left(\frac{\dot{\tilde{\phi}}  \left| m\right| }{\tilde{z}}\right),
\end{align}
\end{subequations}
where the argument of the first kind Bessel's functions $J_n$ is $z$, which is defined just in \eqref{zdef}.
It is worth noting that in this form, it is clear that \eqref{mes3} converges immediately to \eqref{mes2}, when the sources are not considered. 
\\
Integrating \eqref{mes3} two times  we obtain the general family of solutions to the master equation with sources, which reads 
\begin{align}
J_{lm}=&-\frac{iC_1 2^{\frac{1}{2}-2 l} \dot{\tilde{\phi}} ^2 \left| m\right| ^2 z^{-1/2}e^{\frac{1}{2} i (\pi  l+2 z)} \Gamma
	\left(\frac{1}{2}-l\right) \left(-2 z J_{\frac{1}{2}-l}+\overline{S}
	J_{-l-\frac{1}{2}}\right)}{l^2 \left(l^2-1\right)}\nonumber\\
&-\frac{C_22^{2 l+\frac{5}{2}} \dot{\tilde{\phi}} ^2 \left| m\right| ^2 z^{-1/2}e^{-\frac{1}{2} i (\pi  l-2 z)} \Gamma
	\left(l+\frac{3}{2}\right) \left(2 z J_{l-\frac{1}{2}}+\overline{S}
	J_{l+\frac{1}{2}}\right)}{l (l+1)^2 (l+2) }\nonumber\\
&+\int_b^{z}dy\,\int_{a}^{y}d \tilde{z}\, \left(\frac{2^{\frac{1}{2}-2 l} \tilde{z}^{3/2} D_1 e^{\frac{i \pi  l}{2}+i \tilde{z}} \Gamma
	\left(\frac{1}{2}-l\right) \left(K J_{-l-\frac{1}{2}}+L J_{\frac{1}{2}-l}\right)}{(l-1) l}\right.\nonumber\\
&\left.+\frac{i 2^{2
		l+\frac{5}{2}} \tilde{z}^{3/2} D_2 e^{i \tilde{z}-\frac{i \pi  l}{2}} \Gamma \left(l+\frac{3}{2}\right)
	\left(K J_{l+\frac{1}{2}}+L J_{l-\frac{1}{2}}\right)}{(l+1) (l+2)}\right)      \nonumber\\
&+C_3+C_4\frac{\dot{\tilde{\phi}} |m|}{z}.
\label{mes4}
\end{align}
These families of solutions are particularly interesting and useful to explore the dynamics of matter clouds immersed in a Minkowski's background.
\subsection{The Schwarzschild's background}
Now, we show the nonradiative families of solutions, $m=0$, for the vacuum i.e., $G(x)=0$, for equation \eqref{meqsch2}. The solution is expressed in terms of the hypergeometric functions $_2F_1(a_1,a_2;b;z)$, as
\begin{align}
\tilde{J}_{lm}=& (-2)^{-l-2} C_1  M^{-l-2} x^{-l-2} \, _2F_1(2-l,-l;-2 l;2 M x) \nonumber\\
&+ (-2)^{l-1} C_2  M^{l-1} x^{l-1} \, _2F_1(l+1,l+3;2l+2;2 M x).
\label{sch1}
\end{align}
Integrating two times, we find the family of solutions to \eqref{meqsch1}
\begin{align}
J_{lm}=& \frac{C_1 (-1)^{-l} 2^{-l-2} (M x)^{-l} \, _3F_2(-l-1,2-l,-l;1-l,-2 l;2 M x)}{l (l+1) M^2}\nonumber\\
&+\frac{C_2 (-1)^{l+1}
	2^{l-1} x (M x)^l \, _3F_2(l,l+1,l+3;l+2,2 l+2;2 M x)}{l (l+1) M} + C_3x +C_4,
\label{sch2}
\end{align}
where, $_pF_q(a_1,\cdots a_p ;b_1,\cdots, b_q;z)$ are the generalized hypergeometric functions.
\\When we consider the source terms, i.e., $H(x)\ne 0$, the nonradiative solutions to \eqref{meqsch2} reads,
\begin{align}
\tilde{J}_{lm}=&(-1)^{1-l} 2^{-l-2} M^{-l-2} x^{-l-2} \left(A_2 (-1)^{2 l} 2^{2 l+1} M^{2 l+1} x^{2 l+1} \, _2F_1(l+1,l+3;2 l+2;2M x)\right.\nonumber\\
&\left.-A_1 \, _2F_1(2-l,-l;-2 l;2 M x)\right)
 +C_1 (-2)^{-l-2} M^{-l-2} x^{-l-2} \, _2F_1(2-l,-l;-2 l;2 M x)\nonumber\\
&+C_2 (-2)^{l-1} M^{l-1} x^{l-1} \, _2F_1(l+1,l+3;2 l+2;2 M x)
\label{sch3}
\end{align}
where $A_1$, $A_2$ are given by the integrals
\begin{subequations}
\begin{align}
A_1&=-\int_a^xdy\,\frac{(-2)^{l+2} H(y) M^{l+2} y^l \, _2F_1(l+1,l+3;2 (l+1);2 M y)}{B_1+B_2},\\
A_2&=\int_b^x dy\, \frac{(-2)^{1-l} H(y) M^{1-l} y^{-l-1} \, _2F_1(2-l,-l;-2 l;2 M y)}{B_1+B_2},
\end{align}
\end{subequations}
and the functions $B_1$ and $B_2$ are 
\begin{subequations}
\begin{align}
B_1=&(2 M y-1) ((l-2) \, _2F_1(3-l,-l;-2 l;2 M y)\, _2F_1(l+1,l+3;2 (l+1);2 M y),\\
B_2=&\, _2F_1(2-l,-l;-2 l;2 M y) (2 \, _2F_1(l+1,l+3;2 (l+1);2 M y)\nonumber\\
&+(l+1) \,_2F_1(l+2,l+3;2 (l+1);2 M y))).
\end{align}
\end{subequations}
For the radiative ($m\ne 0$) family of solutions to the master equation \eqref{meqsch2} for the vacuum, we find that its most general solution is given by
\begin{align}
\tilde{J}_{lm}=& C_1 L e^{\frac{2\alpha}{xM}} x^{-4}+C_2K \left(2Mx-1\right)^{4\alpha -2}x^{-2-4\alpha} e^{\frac{2\alpha}{xM}},
\label{heunc_sols}
\end{align}
with
\begin{align}
L=H_C\left(-4\alpha ,\beta;\gamma,\delta,\epsilon,\eta\right)\hspace{0.5cm} {\rm and} \hspace{0.5cm}
K=H_C\left(-4\alpha ,-\beta;\gamma,\delta,\epsilon,\eta\right),
\end{align}
where $H_C(\alpha,\beta;\gamma,\delta,\epsilon,\eta)$ are the confluent Heun's functions and their parameters are given by
\begin{subequations}
\begin{align}
\alpha &= i\dot{\tilde{\phi}}m M,\hspace{4.0cm}\beta  = 2-4\alpha\\
\gamma &= 2, \hspace{5.3cm} \delta = 8\alpha(\alpha-1)\\
\epsilon &=-(l+2)(l-1)-8\alpha(\alpha-1)\hspace{0.8cm}\eta =\frac{2Mx-1}{2Mx}.
\end{align}
\label{heun_par}
\end{subequations}
Finally, we present the analytical family of solutions to \eqref{meqsch2} in the radiative case, $m\ne 0$,  when the source terms are considered,
\begin{align}
J_{lm} =&-8M{e^{\frac {2a}{Mx}}} \left( -LMx+{M}^{2}{x}^{2}L+L/4
\right) A_1 x^{-4} \left( 2Mx-1 \right) ^{-2}\nonumber\\
&+2M{e^{\frac {2a}{Mx}}}{x}^{2-4a} \left( 2Mx-1 \right) ^{4a}A_2 K{x}^{-4} \left( 2Mx-1 \right) ^{-2}\nonumber\\
&+C_1 L e^{\frac {2a}{Mx}} x^{-4} + C_2 K e^{\frac {2a}{Mx}} x^{-2-4a}
\left( 2Mx-1 \right)^{-2+4a},
\end{align}
where $A_1$ and $A_2$ are the integrals
\begin{subequations}
\begin{align}
A_1=&\int_a^x d\tilde{x}\, \frac{\tilde{x}^2 H(\tilde{x}) e^{-\frac{2a}{M\tilde{x}}}K}{-4\,
	LKM\tilde{x}+8\,LKaM\tilde{x}-LS+2\,LM\tilde{x}S+KR-2\,KM\tilde{x}R} \label{a1}\\
A_2=&\int_b^x d\tilde{x}\, \frac{4\tilde{x}^{4a}{e^{-\frac {2a}{M\tilde{x}}}}H(\tilde{x})
\left( M\tilde{x}-1/2 \right) ^{2} \left( 2M\tilde{x}-1 \right) ^{-4a}L}{-4
\,LKM\tilde{x}+8\,LKaM\tilde{x}-LS+2\,LM\tilde{x}S+KR-2\,KM\tilde{x}R}\label{a2} ,
\end{align}
\end{subequations}
where $S$ and $R$ are the derivative of the Heun's functions, i.e. $S=K'(x)$ and $R=L'(x)$, in which we suppress all indices except one which gives the functional dependence. 
\section{Families of solutions for $l=2$}
Now, we show that the families of solutions found here are reduced to those previously reported in the literature for $l=2$. Thus, for this particular value of $l$ we obtain that the family of solutions to the master equation for the vacuum, \eqref{field_eq_x13} takes the explicit form
\begin{equation}
\tilde{J}_{lm}=E_1 x+\frac{E_2 e^{\frac{2 i \dot{\tilde{\phi }} \left| m\right| }{x}} \left(6 x^3 \dot{\tilde{\phi }} \left| m\right| -6 i x^2 \dot{\tilde{\phi }}^2 \left| m\right| ^2-4 x
	\dot{\tilde{\phi }}^3 \left| m\right| ^3+2 i \dot{\tilde{\phi }}^4 \left| m\right| ^4+3 i x^4\right)}{4 x^3 \dot{\tilde{\phi }}^5 \left| m\right| ^5}
\label{fam1}
\end{equation}
Now, substituting $l=2$ in the family of solutions \eqref{mes2}, one obtains
\begin{align}
\tilde{J}_{lm}=&\frac{i C_1 \dot{\tilde{\phi}} ^3 \left| m\right| ^3 e^{\frac{2 i \dot{\tilde{\phi}}  \left| m\right| }{x}}}{6 x^3}-\frac{40 i C_2 \dot{\tilde{\phi}} ^3 \left| m\right| ^3 e^{\frac{2 i \dot{\tilde{\phi}}  \left| m\right|
		}{x}}}{x^3}-\frac{C_1 \dot{\tilde{\phi}} ^2 \left| m\right| ^2 e^{\frac{2 i \dot{\tilde{\phi}}  \left| m\right| }{x}}}{3 x^2} +\frac{80 C_2 \dot{\tilde{\phi}} ^2 \left| m\right| ^2 e^{\frac{2 i \dot{\tilde{\phi}}  \left| m\right|
	}{x}}}{x^2} \nonumber
\\&-\frac{i C_1 \dot{\tilde{\phi}}  \left| m\right|  e^{\frac{2 i \dot{\tilde{\phi}}  \left| m\right| }{x}}}{2 x}+\frac{120 i C_2 \dot{\tilde{\phi}}  \left| m\right|  e^{\frac{2 i \dot{\tilde{\phi}}  \left| m\right|
}{x}}}{x}+\frac{1}{2} C_1 e^{\frac{2 i \dot{\tilde{\phi}}  \left| m\right| }{x}} -120 C_2 e^{\frac{2 i \dot{\tilde{\phi}}  \left| m\right| }{x}}\nonumber
\\&+\frac{i C_1 x e^{\frac{2 i \dot{\tilde{\phi}}  \left| m\right| }{x}}}{4 \dot{\tilde{\phi}} 
\left| m\right| }+\frac{i C_1 x}{4 \dot{\tilde{\phi}}  \left| m\right| }-\frac{60 i C_2 x e^{\frac{2 i \dot{\tilde{\phi}}  \left| m\right| }{x}}}{\dot{\tilde{\phi}}  \left| m\right| }+\frac{60 i  x}{\dot{\tilde{\phi}}  \left|
m\right| }.
\label{fam2}
\end{align} 
Both family of solutions, \eqref{fam1} and \eqref{fam2}, are completely equivalent. Note that, the transformation between the constants, necessary to pass from \eqref{fam1} to \eqref{fam2} is given by
\begin{equation}
E_1= \frac{i \left(C_1+240 C_2\right)}{4 \dot{\tilde{\phi}}  \left| m\right| },\hspace{1cm}
E_2= \frac{1}{3} \left(C_1-240 C_2\right) \dot{\tilde{\phi}} ^4 \left| m\right| ^4.
\end{equation}
Note that for the Schwarzschild case, when no sources are present, the master equation \eqref{meqsch2}, for the vacuum and $l=2$ takes the explicit form
\begin{align}
x^2 (2 M x-1) \tilde{J}_{lm,xx}+2 x (7 M x-2) \tilde{J}_{lm,x}+ (16 M x+4)\tilde{J}_{lm}=0.
\end{align}
Its family of solutions is
\begin{equation}
\tilde{J}_{lm}=\frac{C_1}{x^4}-\frac{C_2 \left(16 M^4 x^4+32 M^3 x^3-44 M^2 x^2-4 M x+12 (1-2 M x)^2 \log (1-2 M x)+7\right)}{64
	M^5 x^4 (1-2 M x)^2}.
\end{equation}
Now, specializing the solutions \eqref{sch1} for $l=2$, we find a totally equivalent solution, i.e.,
\begin{equation}
\tilde{J}_{lm}=\frac{D_1}{16 M^4 x^4}+\frac{5 D_2 \left(2 M x \left(2 M^3 x^3+4 M^2 x^2-9 M x+3\right)+3 (1-2 M x)^2 \log (1-2 M
	x)\right)}{8 M^4 x^4 (1-2 M x)^2}.
\end{equation}
Thus, a simple Maclaurin series expansion of both solutions shows that the relationship between the constants is
\begin{align}
D_1= \frac{64 C_1 M^5-7 C_2}{4 M},\hspace{1cm} D_2 = -\frac{C_2}{10 M}.
\end{align}
Finally, given that the known family of solutions for $l=2$ is written in terms of power of series around the point $r=2M$, as shown in \cite{B05}, we expand the radiative family of solutions for the master equation \eqref{meqsch1} around the same point $r=2M$ for $l=2$. Thus, we observe that the confluent Heun's function $H_C(-4\alpha,\beta;\gamma,\delta,\epsilon,\eta)$ is expressed as a Taylor series for the parameters \eqref{heun_par} around $\eta=0$, namely
\begin{align}
H_C(-4\alpha,\beta;\gamma,\delta,\epsilon,\eta)\simeq& 1+{\frac { \left( (4a+1)^2-5+(l-1)(l+2) \right) \eta}{-3+4\,a}}\nonumber\\
&+\frac{1}{8 (a-1) (4 a-3)}
\left( \left(256 a^4+192 a^3+32 a^2 \left(l^2+l-5\right) \right.\right.\nonumber\\
&+\left.\left. 4 a \left(4 l^2+4 l-39\right)+l^4+2 l^3-17 l^2-18 l+72\right)\eta^2\right),
\label{Heun1}
\end{align}
and for the confluent Heun's function $H_C(-4\alpha,-\beta;\gamma,\delta,\epsilon,\eta)$, 
\begin{align}
H_C(-4\alpha,-\beta;\gamma,\delta,\epsilon,\eta)\simeq&1-\frac{\left(4 a+l^2+l\right)\eta}{4 a-1}-\frac{\left(12 a-l^4-2 l^3+l^2+2 l\right)\eta^2 }{8 a (4 a-1)}.
\label{Heun2}
\end{align}
Then, from \eqref{Heun1} and \eqref{Heun2} we obtain that around to $r=2M$, \eqref{heunc_sols} at first order for $l=2$, 
\begin{align}
\tilde{J}_{lm}=& C_1 \left(\frac{16 e^{4 \alpha } (4 \alpha +12) \eta  M^4}{4 \alpha -3}+16 e^{4 \alpha } M^4\right)
-\frac{2^{4 \alpha +2} C_2 e^{4 \alpha } \left(16 \alpha ^2+16 \alpha +2\right) \eta ^{4 \alpha -1}	\left(\frac{1}{M}\right)^{-4 \alpha -2}}{4 \alpha -1}\nonumber
\\
&+\frac{2^{4 \alpha -1} C_2 e^{4 \alpha } \left(256 \alpha^4+576 \alpha ^3+384 \alpha ^2+132 \alpha +24\right) \eta ^{4 \alpha } \left(\frac{1}{M}\right)^{-4 \alpha	-2}}{\alpha  (4 \alpha -1)}\nonumber\\
&-\frac{2^{4 \alpha +1} C_2e^{4 \alpha } \left(256 \alpha ^5+896 \alpha ^4+1056 \alpha^3+636 \alpha ^2+228 \alpha +72\right) \eta ^{4 \alpha +1} \left(\frac{1}{M}\right)^{-4 \alpha -2}}{3 \alpha(4 \alpha -1)}\nonumber\\
&+2^{4 \alpha +2} C_2 e^{4 \alpha } \eta ^{4 \alpha -2} \left(\frac{1}{M}\right)^{-4 \alpha -2},
\end{align}
that are just the family of solutions for the master equation obtained using power series around $r=2M$.
\section{Summary and Conclusions}
In this work we report new solutions to the master equation when a flat background is considered, generalizing the results obtained by M\"adler \cite{M13} with the inclusion of source terms. Likewise, we reexpress the family of solutions for the vacuum using only Bessel's functions of the first kind. 
\\Bishop \cite{B05} already found the solutions to the field equations in the space-time exterior to a static and spherically symmetric black-hole, for $l=2$, but only by expanding the metric variables in power series around the coordinate singularity $r=2M$, and in an asymptotic expansion near the null infinity. However his solutions depend on the order of the expansion and in this sense it is an approximation. We report for the first time in the literature the exact solutions to the master equation in terms of the hypergeometric (Heun's function) for the nonradiative (radiative) modes with and without source terms. Considering the solutions for $l=2$ we also show the equivalence between our solution and those reported in the literature. 
\\
Finally, notice that the importance of these analytical results is in the fact that it can be useful in the construction of semianalytical models for matter distributions for this regime, like thin and thick shells or stars composed of layers obeying some equation of state.   However, as already mentioned, it is important to bear in mind that the matter fields must
be known {\it a priori} throughout the spacetime.
\begin{acknowledgments}
We would like to thank the Brazilian agencies CAPES, FAPESP (2013/11990-1) and CNPq (308983/2013-0) by the financial support. 
We would also like to thank the referees for their helpful comments and suggestions.
\end{acknowledgments}

\providecommand{\noopsort}[1]{}\providecommand{\singleletter}[1]{#1}%

\end{document}